\documentclass[a4paper,12pt]{article}

\usepackage{amssymb,amsmath}

\title{Superintegrability in a non-conformally-flat space}
\author{%
E.~G.~Kalnins\\Department of Mathematics, University of Waikato,\\Hamilon, New Zealand\\[5mm]%
J.~M.~Kress\\School of Mathematics and Statistics, University if New South Wales,\\ Sydney, Australia\\[5mm]%
W.~Miller,~Jr.\\School of Mathematics, University of Minnesota,\\ Minneapolis, Minnesota, U.~S.~A.%
}
 
\begin{document}

\maketitle

\begin{abstract}
Superintegrable systems in two- and three-dimensional
spaces of constant curvature have been extensively studied.  From
these, superintegrable systems in conformally flat spaces can be
constructed by St\"ackel transform.  In this paper a method
developed to establish the superintegrability of the Tremblay-Turbiner-Winternitz system
in two dimensions is
extended to higher dimensions and a superintegrable system on a
non-conformally-flat four-dimensional space is found.  In doing so,
curvature corrections to the corresponding classical potential are
found to be necessary.  It is found that some subalgebras of the symmetry algebra close polynomially.
\end{abstract}

A maximally superintegrable quantum system on a $n$-dimensional manifold is an integrable
Hamiltonian system of $n$ mutually commuting differential operators and an additional
$n-1$ 
differential operators so that the full $2n-1$ are algebraically independent and commute
with one distinguished operator, the Hamiltonian, which we will take to have the form 
$H=\nabla^2+V$.  
In all previously known quantun superintegrable systems of this
form with non-constant potential, $\nabla^2$ is the
the natural Laplacian of a constant curvature manifold.  
A St\"ackel transform can be
used to construct systems on conformally-flat manifolds from systems of constant
curvature manifolds \cite{V,CCM}, but no systems have been previously exhibited on a
non-conformally-flat manifold.

In \cite{KKM2010} a classical superintegrable
system on a non-conformally flat 4-dimensional space was found by generalising
the two-dimensional Tremblay-Turbiner-Winternitz (TTW) system
\cite{TTW} and here we will show that this system can be quantised in a way that
preserves its superintegrability.  

The two-dimensional TTW system is integrable by virtue of a second order operator
associated with its separability.  Its
superintegrability has been demonstrated by constructing
a symmetry from raising and lowering operators built out of special function
recurrence relations that act on an eigenbasis of separated solutions \cite{KKM2011}. 
A similar approach has been used to generate other families of superintegrable systems
in two dimensions \cite{KKM2011,PTV2012,PVZ2011,M2010} and in so doing, greatly expanded the
list of superintegrable systems
with higher order symmetries.   
% Recently, an example of non-separable quantum superintegrable
% systems has been found \cite{PW2011}.  
Previous studies of higher order superintegrability have uncovered quantum superintegrable systems with
no classical counterpart \cite{GW2002} and the need to consider systems in higher dimensions
with higher order symmetries
has been highlighted recently by the use of higher order symmetries to determine the spectrum
of a deformed Kepler-Coulomb system in three dimensions \cite{TD2011}.

Here we extend the raising and lowering operator method used on the quantum TTW system to
higher
dimensions and construct sufficient additional algebraically independent operators to
show that it is 
superintegrable.  Furthermore, we find that some subalgebras of symmetry operators close
polynomially as is common with previously known superintegrable systems.

An interesting feature encountered below is that in order to construct the additional
symmetries required for superintegrability the potential must be deformed by the addition
of curvature
terms that make the Hamiltonian conformally covariant.  These terms are not simply
the usual minimal choice for a conformally covariant Laplacian, namely, $-\mathcal{R}/6$,
where $\mathcal{R}$ is the scalar curvature associated with the underlying metric, but
also include an invariant constructed from the Weyl conformal curvature.

The system considered below is 4-dimensional only so as to provide the simplest
non-conformally-flat example.  The procedure can be extended to higher dimensions with no
greater difficulty and a number of other systems similar to the TTW system
\cite{KKM2011} could also be
extended.

\section{The classical 4D non-conformally-flat system}

The Tremblay-Turbiner-Winternitz (TTW) system \cite{TTW} sparked great interest because it
provided an infinite
family of superintegrable systems and examples of systems with arbitrarily high degree 
 symmetries. The system in polar coordinates is given by
\[
 H_{TTW} = p_r^2 + \alpha r^2 + \frac1{r^2}\left( p_\theta^2 +
\frac{\beta_1}{\cos^2(k\theta)}
 + \frac{\beta_2}{\sin^2(k\theta)}\right).
\]
The superintegrability of both classical \cite{KMP2011} and quantum versions
\cite{KKM2009}
were established for all positive rational values of the parameter $k$ along with the
polynomial closure of its symmetry algebra and the general approach was quickly
extended to other families of systems
in two dimensions such as
\[
 H = \cosh^2\psi\left(p_\psi^2+p_\varphi^2 + \frac{\alpha}{\cos^2k\varphi}
 + \frac{\beta}{\sin^2k\varphi} + \frac{\gamma}{\sinh^2\psi}\right).
\]
In the classical case, this approach as also been extended to higher dimensions and the
 4-dimensional generalisation of the classical TTW system,
\begin{eqnarray}
\label{eqn4DTTW}
H = L_1 &=& p_r^2 + \alpha r^2 + \frac{L_2}{r^2} \\
L_2 &=& p_{\theta_1}^2 + \frac{\beta_1}{\cos^2(k_1\theta_1)} +
\frac{L_3}{\sin^2(k_1\theta_1)} \nonumber \\
L_3 &=& p_{\theta_2}^2 + \frac{\beta_2}{\cos^2(k_2\theta_2)} +
\frac{L_4}{\sin^2(k_2\theta_2)}  \nonumber \\
L_4 &=& p_{\theta_3}^2 + \frac{\beta_3}{\cos^2(k_3\theta_3)} +
\frac{\beta_4}{\sin^2(k_3\theta_3)}, \nonumber 
\end{eqnarray}
was shown to be superintegrable for all positive rational $k_1$, $k_2$ and $k_3$
\cite{KKM2010}. Here, 
the underlying manifold, on which this is a natural Hamiltonian system, has metric
\begin{equation}
\label{metric}
 g = e_0\otimes e_0 + e_1\otimes e_1 + e_2\otimes e_2 + e_3\otimes e_3,
\end{equation}
\[
e_0 = dr,\quad 
e_1 = rd\theta_1,\quad 
e_2 = r\sin(k_1\theta_1)d\theta_2,\quad 
e_3 = r\sin(k_1\theta_1)\sin(k_2\theta_2)d\theta_3,
\]
and is no longer flat unless 
$k_1=k_2=1$.  Furthermore, in these coordinates, each component of the Weyl conformal
tensor is a
constant multiple of
\[
 \frac{k_1^2-k_2^2}{r^2\sin^2(k_1\theta_1)}
\]
and so the underlying manifold is only conformally flat when this quantity
vanishes, that is, $k_1=k_2$. In \cite{KKM2010} it was found that
(\ref{eqn4DTTW}) is superintegrable with the 4 second order constants given above as well
as an additional cubic and
two quartic constants.

A natural question to ask is whether there is a corresponding quantum system with the same
or minimally modified potential.
It is straightforward to check that 
\begin{eqnarray*}
 H  &=& \partial_r^2+\frac{3}{r}\partial_r - \omega^2 r^2 + \frac{L_1}{r^2}  \\ 
 %+\frac{1-k_1^2}{r^2} \\
%  &=&  \nabla^2  + V_0 + \frac{1-k_1^2}{r^2} +
% \frac{k_1^2-k_2^2}{4r^2\sin^2(k_1\theta_1)}
%  \ = \  \nabla^2  + V_0 - \frac16{\mathcal R} - \frac1{24}{\mathcal W}
 L_1 &=& \partial_{\theta_1}^2 + 2k_1\cot(k_1\theta_1)\partial_{\theta_1} +
\frac{\beta_1}{\cos^2(k_1\theta_1)}
                + \frac{L_2}{\sin^2(k_1\theta_1)} \\
%                 + \frac{k_1^2-k_2^2}{4\sin^2(k_1\theta_1)} \\
 L_2 &=& \partial_{\theta_2}^2 + k_2\cot(k_2\theta_2)\partial_{\theta_2} +
\frac{\beta_2}{\cos^2(k_2\theta_2)}
                + \frac{L_3}{\sin^2(k_2\theta_2)} \\
 L_3 &=& \partial_{\theta_3}^2 + \frac{\beta_3}{\cos^2(k_3\theta_3)} +
\frac{\beta_4}{\sin^2(k_3\theta_3)}
\end{eqnarray*}
are four mutually commuting differential operators and  $H$ is
a Hamiltonian of the form $H=\nabla^2+V_0$ where $\nabla^2$ is the Laplacian on 
a 4-dimensional manifold with metric (\ref{metric}) and
\[
 V_0 = \alpha r^2 + \frac{\beta_1}{r^2\cos^2(k_1\theta_1)} +
\frac{\beta_2}{r^2\sin^2(k_1\theta_1)\cos^2(k_2\theta_2)}
 + \frac{\beta_3}{r^2\sin^2(k_1\theta_1)\sin^2(k_2\theta_2)\cos^2(k_3\theta_3)}
\]
\begin{equation}
 \label{V0}
 {} + \frac{\beta_4}{r^2\sin^2(k_1\theta_1)\sin^2(k_2\theta_2)\sin^2(k_3\theta_3)}.
\end{equation}
However, it is not a simple matter to quantise the additional classical constants found in
\cite{KKM2010}, and
so to investigate whether this system remains superintegrable in the quantum case we attempt to adapt the raising and
lowering operator methods from \cite{KKM2011}.

In the two-dimensional examples, with parameter $k$, 
the essence of the method is that solutions can be found by separation of variables with the separated eigenfunctions
enumerated by positive integers $n_0$ and $n_1$.  The energy eigenvalue of each separated
solution depended only on the combination
$n_0+kn_1$.  Differential operators were then constructed to raise or lower $n_0$ or $n_1$ by integer amounts and so that for 
rational $k$, compositions of these operators could be found that left $n_0+k_1n_1$
unchanged and hence preserved the energy. While these additional operators were
constructed to act on a separated eigenbasis, they
were found to in fact be expressible as differential operators. It should be noted here
that the linear dependence of the energy on the quantum numbers $n_0$ and $n_1$ is crucial
to the method and maintaining this below leads to the need for quantum corrections to the
potential.

In order to extended this approach to our present example we must first 
solve the system by separation of variables and so we postulate a solution to $H\Psi =
E\Psi$ of the form
\[
\Psi = \Psi_0(r)\Psi_1(\theta_1)\Psi_2(\theta_2)\Psi_3(\theta_3),
\]
with
\[
L_3\Psi_3=\ell_3\Psi_3,\qquad
L_2\Psi_2\Psi_3=\ell_2\Psi_2\Psi_3,\qquad
L_1\Psi_1\Psi_2\Psi_3=\ell_1\Psi_1\Psi_2\Psi_3.
\]
While the solution of the separated equations is unremarkable, the details
are written out at length so as to expose the point at which the quantum corrections
(\ref{eqnQcorrection1}) and (\ref{eqnQcorrection2}) to the potential become necessary.

We find that each angular equation is, up to a gauge scaling, of the form
\[
u''(y)+\left(\frac{\frac14-\alpha^2}{\sin^2y}+\frac{\frac14-\beta^2}{\cos^2y}
   + (2n+\alpha+\beta+1)^2\right)u(y) = 0,
\]
which has solution
\[
u(y) = (\sin y)^{\alpha+\frac12}(\cos y)^{\beta+\frac12} P_n^{(\alpha,\beta)}(\cos 2y)
\]
where $P_n^{(\alpha,\beta)}(x)$ is a Jacobi function \cite{MOS}.

Starting with the $\theta_3$ equation, we make the replacements
\[
 \beta_3=k_3^2\left(\frac14-a_4^2\right),\qquad
 \beta_4=k_3^2\left(\frac14-a_3^2\right),
\]
and find the separated equation is
\[
 L_3\Psi_3(\theta_3) = \Psi_3''(\theta_3) 
                + \left(\frac{k_3^2\left(\frac14-a_4^2\right)}{\cos^2(k_3\theta_3)}
                + \frac{k_3^2\left(\frac14-a_3^2\right)}{\sin^2(k_3\theta_3)}\right)\Psi_3(\theta_3) = \ell_3\Psi_3(\theta_3)
\]
which has solutions
\[
 \Psi_{3,n_3}^{a_3,a_4}(\theta_3)=(\sin(k_3\theta_3))^{a_3+\frac12}(\cos(k_3\theta_3))^{a_4+\frac12}P^{(a_3,a_4)}_{n_3}(\cos(2k_3\theta_3))
\]
with eigenvalues
\begin{equation}
\label{eqnl3}
 L_3\Psi_{3,n_3}^{a_3,a_4}(\theta_3) = \ell_3\Psi_{3,n_3}^{a_3,a_4}(\theta_3),\qquad
 \ell_3 = -k_3^2(2n_3+a_3+a_4+1)^2.
\end{equation}
The separated equation $L_2\Psi_2(\theta_2) = \ell_2\Psi_2(\theta_2)$ is now 
\[
 \Psi_2''(\theta_2) + k_2\cot(k_2\theta_2)\Psi_2'(\theta_2)
 + \left(\frac{\beta_2}{\cos^2(k_2\theta_2)}
                + \frac{\ell_3}{\sin^2(k_2\theta_2)}\right)\Psi_2(\theta_2) = \ell_2\Psi_2(\theta_2)
\]
which we transform with
\[
\Psi_2(\theta_2) = (\sin(k_2\theta_2))^{-\frac12}\psi_2(\theta_2)
\]
to absorb the first derivative term to give
\[
\psi_2''(\theta_2)
 + \left(\frac{\beta_2}{\cos^2(k_2\theta_2)}
       + \frac{\ell_3+\frac14k_2^2}{\sin^2(k_2\theta_2)}
       + \frac14k_2^2-\ell_2\right)\psi_2(\theta_2) = 0
\]
and we make the replacements
\[
\beta_2=k_2^2\left(\frac14-a_2^2\right),\qquad
\ell_3+\frac14k_2^2 = k_2^2\left(\frac14-A_2^2\right),
\]
which when combined with (\ref{eqnl3}) gives
\begin{equation}
\label{eqnA2}
A_2 = \frac{k_3}{k_2}(2n_3+a_3+a_4+1). 
\end{equation}
The separated $\theta_2$ equation becomes
\[
\psi_2''(\theta_2)
 + \left(\frac{k_2^2\left(\frac14-a_2^2\right)}{\cos^2(k_2\theta_2)}
       + \frac{k_2^2\left(\frac14-A_2^2\right)}{\sin^2(k_2\theta_2)}
       + \frac{k_2^2}4-\ell_2\right)\psi_2(\theta_2) = 0
\]
where
\begin{equation}
\label{eqnl2}
\frac{k_2^2}4-\ell_2   = k_2^2(2{n_2}+a_2+A_2+1)^2
\end{equation}
and has solution
\[
\Psi_{2,n_2}^{A_2,a_2}(\theta_2) = (\sin(k_2\theta_2))^{A_2}(\cos(k_2\theta_2))^{a_2+\frac12}
   P_{n_2}^{(A_2,a_2)}(\cos(2k_2\theta_2)).
\]
The separated $\theta_1$ equation $L_1\Psi_1(\theta_1) = \ell_1\Psi_1(\theta_1)$ is
\begin{equation}
\label{eqntheta1}
  \Psi_1''(\theta_1) + 2k_1\cot(k_1\theta_1)\Psi_1'(\theta_1)
 + \left(\frac{\beta_1}{\cos^2(k_1\theta_1)}
                + \frac{\ell_2}{\sin^2(k_1\theta_1)}\right)\Psi_1(\theta_1) = \ell_1\Psi_1(\theta_1),
\end{equation}
which we transform with
\[
\Psi_1(\theta_1) = (\sin(k_1\theta_1))^{-1}\psi(\theta_1)
\]
to absorb the first order term to give
\begin{equation}
\label{eqntheta1g}
\psi_1''(\theta_1)
 + \left(\frac{\beta_1}{\cos^2(k_1\theta_1)}
       + \frac{\ell_2}{\sin^2(k_1\theta_1)}
       + k_1^2-\ell_1\right)\psi_1(\theta_1) = 0
\end{equation}
and we make the replacements
\[
\beta_1=k_1^2\left(\frac14-a_1^2\right),\qquad
\ell_2 = k_1^2\left(\frac14-A_1^2\right).
\]
Combining this with (\ref{eqnl2}) gives
\[
A_1 = \sqrt{\frac14\left(1-\frac{k_2^2}{k_1^2}\right) + \frac{k_2^2}{k_1^2}(2n_2+a_2+A_2+1)^2}.
\]
This does not have the same form as (\ref{eqnA2}) and will not lead to an energy eigenvalue
that depends linearly on $n_2$.  Hence, we instead propose an additional quantum correction in the potential of 
\begin{equation}
 \label{eqnQcorrection1}
\hat V_1 =  \frac{\frac14(k_1^2-k^2)}{r^2\sin^2(k_1\theta_1)},
\end{equation}
which in turn leads to a modified (\ref{eqntheta1g}),  
\[
\psi_1''(\theta_1)
 + \left(\frac{\beta_1}{\cos^2(k_1\theta_1)}
       + \frac{\ell_2+\frac14(k_1^2-k_2^2)}{\sin^2(k_1\theta_1)}
       + k_1^2-\ell_1\right)\psi_1(\theta_1) = 0.
\]
Now, making the replacements
\[
\beta_1=k_1^2\left(\frac14-a_1^2\right),\qquad
\ell_2 + \frac14(k_1^2-k_2^2)= k_1^2\left(\frac14-A_1^2\right)
\]
gives
\begin{equation}
\label{eqnA1}
A_1 = \frac{k_2}{k_1}(2n_2+A_2+a_2+1).
\end{equation}
The separated $\theta_1$ equation is now
\[
\psi_1''(\theta_1)
 + \left(\frac{k_1^2\left(\frac14-a_1^2\right)}{\cos^2(k_1\theta_1)}
       + \frac{k_1^2\left(\frac14-A_1^2\right)}{\sin^2(k_1\theta_1)}
       + k_1^2-\ell_1\right)\psi_1(\theta_1) = 0
\]
where
\[
k_1^2-\ell_1 = k_1^2(2n_1+A_1+a_1+1)^2
\]
and has solutions
\[
\Psi_{1,n_1}^{A_1,a_1}(\theta_1) = (\sin(k_1\theta_1))^{A_1-\frac12}(\cos(k_1\theta_1))^{a_1+\frac12}
   P_{n_1}^{(A_1,a_1)}(\cos(2k_1\theta_1)).
\]
Finally, the separated radial equation is 
\begin{equation}
\label{eqnr}
 H\Psi_0(r) = \partial_r^2\Psi_0(r)+\frac{3}{r}\partial_r\Psi_0(r)
 + \left(-\omega^2 r^2 + \frac{\ell_1}{r^2}\right)\Psi_0(r) = E\Psi_0(r).
\end{equation}
In a similar way to above, in order that $E$ depend linearly on $n_1$, we propose the addition of a quantum correction
to the potential of
\begin{equation}
 \label{eqnQcorrection2}
\hat V_2 = \frac{1-k_1^2}{r^2}
\end{equation}
which leads to a modified version of (\ref{eqnr}),
\[
 H\Psi_0(r) = \partial_r^2\Psi_0(r)+\frac{3}{r}\partial_r\Psi_0(r)
 + \left(-\omega^2 r^2 + \frac{\ell_1-k_1^2+1}{r^2}\right)\Psi_0(r) = E\Psi_0(r).
\]
We remove the first order terms with the transformation
\[
\Psi_0(r) = r^{-\frac32}\psi_0(r)
\]
to give
\[
\partial_r^2\psi_0(r) + \left(-\omega^2 r^2 + \frac{\frac14-k_1^2+\ell_1}{r^2} - E\right)\psi_0(r) = 0.
\]
Now,
\[
u''(x) + \left(-x^2+\frac{\frac14-A_0^2}{x^2}+4n+2A_0+2\right)u(x) = 0
\]
has solution
\[
u(x) = e^{-\frac{x^2}2}x^{A_0+\frac12}L_n^{(A_0)}(x^2),
\]
where $L_n^{(A_0)}(x)$ is a Laguerre function \cite{MOS}.

We needed $ A_0^2=k_1^2-\ell_1$
and we already have
$
k_1^2-\ell_1 = k_1^2(2n_1+A_1+a_1+1)^2
$
so
\begin{equation}
\label{eqnPsi0}
\Psi_{0,n_0}^{A_0}(r) = \omega^{A_0/2} e^{-\frac{\omega r^2}2}r^{A_0-1}L_{n_0}^{(A_0)}(\omega r^2),
\end{equation}
where the mulitplicative factor of $\omega^{A_0/2}$ is chosen for later convenience, and
\begin{equation}
\label{eqnA0}
A_0 = k_1(2n_1+a_1+A_1+1),\qquad
E = -\omega(4n_0+2A_0+2).
\end{equation}
Note that with the quantum deformation (\ref{eqnQcorrection2}) the relationship of $A_0$ to $n_1$ is
similar to that seen in (\ref{eqnA2}) and (\ref{eqnA1}). 

Now, putting together (\ref{eqnA2}), (\ref{eqnA1}) and (\ref{eqnA0}) we find
\begin{equation}
\label{eqnE}
E = -2\omega(2n_0 + 2k_1n_1 + 2k_2n_2 + 2k_3n_3 + k_1a_1 + k_2a_2 + k_3a_3 + k_3a_4 + k_1 + k_2 + k_3 + 1)
\end{equation}
for a solution of the form
\[
 \Psi_{n_0,n_1,n_2,n_3} = \Psi_{0,n_0}^{A_0}(r)\Psi_{1,n_1}^{A_1,a_1}(\theta_1)\Psi_{2,n_2}^{A_2,a_2}(\theta_2)\Psi_{3,n_3}^{a_3,a_4}(\theta_3).
\]

With the quantum corrections (\ref{eqnQcorrection1}) and (\ref{eqnQcorrection2}) 
added to the potential we have the following set of mutually commuting differential
operators.
\begin{eqnarray*}
 H \ = \ L_0 &=& \partial_r^2+\frac{3}{r}\partial_r - \omega^2 r^2 + \frac{L_1}{r^2}
 + \frac{1-k_1^2}{r^2} \\
%  &=&  \nabla^2  + V_0 + \frac{1-k_1^2}{r^2} + \frac{k_1^2-k_2^2}{4r^2\sin^2(k_1\theta_1)}
%  \ = \  \nabla^2  + V_0 - \frac16{\mathcal R} - \frac1{24}{\mathcal W}
 L_1 &=& \partial_{\theta_1}^2 + 2k_1\cot(k_1\theta_1)\partial_{\theta_1} + \frac{\beta_1}{\cos^2(k_1\theta_1)}
                + \frac{L_2}{\sin^2(k_1\theta_1)}
                + \frac{k_1^2-k_2^2}{4\sin^2(k_1\theta_1)} \\
 L_2 &=& \partial_{\theta_2}^2 + k_2\cot(k_2\theta_2)\partial_{\theta_2} + \frac{\beta_2}{\cos^2(k_2\theta_2)}
                + \frac{L_3}{\sin^2(k_2\theta_2)} \\
 L_3 &=& \partial_{\theta_3}^2 + \frac{\beta_3}{\cos^2(k_3\theta_3)} +
\frac{\beta_4}{\sin^2(k_3\theta_3)}. \\
\end{eqnarray*}
$H\Psi = E\Psi$ remains separable with the additional terms.  In the following, we use
these redefined $H$ and $L_1$.

For the metric (\ref{metric}), the scalar curvature is
\[
 \mathcal{R} = -\frac{6}{r^2} + k_1^2\left(\frac6{r^2}-\frac{2}{r^2\sin^2(k_1\theta_1)}\right)
    + \frac{2k_2^2}{r^2\sin^2(k_1\theta_1)},
\]
and with Weyl conformal tensor $W_{abcd}$, if we define
\[
\mathcal{W} =  \sqrt{3W_{abcd}W^{abcd}} = \frac{2(k_1^2-k_2^2)}{r^2\sin^2(k_1\theta_1)}. 
\]
then
\[
 H \ = \  \nabla^2  + V_0 + \hat V_1 + \hat V_2 \ = \  \nabla^2  + V_0 - \frac16{\mathcal
R} - \frac1{24}{\mathcal W}.
\]
Note that $\nabla^2  + \hat V_1 + \hat V_2$ is a conformally covariant 
Laplacian and the 
metric $g$ is conformally flat if and only if $k_1=k_2$.

\section{Raising and lowering operators}

Our aim is now to use special function identities to raise and lower the
$n_i$ while preserving $E$ and
produce new operators commuting with $H$.

Using differential identities for Laguerre functions \cite{MOS} we
construct the operators that act on the radial part of the separated
solutions,
\begin{eqnarray*}
 K^{+\ A_0}_{0\ n_0} &=& \frac{1-A_0}r\partial_r + (2n_0+A_0+1)\omega
+ \frac{1-A_0^2}{r^2}, \\
 K^{-\ A_0}_{0\ n_0} &=& \frac{1+A_0}r\partial_r + (2n_0+A_0+1)\omega
+ \frac{1-A_0^2}{r^2}.
\end{eqnarray*}
These raise or lower $n_0$ by $1$ while simultaneously lowering or
raising and $A_0$ by 2, that is,
\begin{eqnarray*}
 K^{+\ A_0}_{0\ n_0}\Psi^{A_0}_{n_0} &=&
-2\omega(n_0+1)(n_0+A_0)\Psi^{A_0-2}_{n_0+1}, \\
 K^{-\ A_0}_{0\ n_0}\Psi^{A_0}_{n_0} &=&
-2\omega\Psi^{A_0+2}_{n_0-1}. 
\end{eqnarray*}
Note that the constant multiplicative factor of $\omega^{A_0/2}$ in (\ref{eqnPsi0}) was chosen so
that both of these have a factor of $\omega$ on the right hand side.

For the angular functions, we can use Jacobi function identities
\cite{MOS} to make operators
that raise and lower $n$ alone,
\begin{eqnarray*}
J_{n}^{+} &=& -\frac{(N+1)\sin(2k\theta)}{2k}\partial_{\theta}
  - \frac12\Bigl((N+1)(N+1-c-d)\cos(2k\theta) \\
 & & {} \quad -(N+1)(c-d)+a^2-b^2\Bigr), \\
J_{n}^{-} &=& \frac{(N-1)\sin(2k\theta)}{2k}\partial_{\theta} 
            - \frac12\Bigl((N-1)(N-1+c+d)\cos(2k\theta) \\
 & & {} \quad  + (N-1)(c-d) + a^2-b^2\Bigr),
\end{eqnarray*}
where 
$N = 2n+a+b+1$ and their action on
\[
\Theta^{(a,b)}_n =
\sin^{a+c}(k\theta)\cos^{b+d}(k\theta)P_{n}^{(a,b)}\cos(2k\theta)
\]
is given by
\begin{eqnarray*}
J_{n}^{+}\Theta^{(a,b)}_n
       &=&-2({n}+1)({n}+a+b+1)\Theta^{(a,b)}_{n+1}, \\
J_{n}^{-}\Theta^{(a,b)}_n
       &=&-2({n}+a)({n}+b)\Theta^{(a,b)}_{n-1}.
\end{eqnarray*}
The operators above are essentially those used in \cite{KKM2011} and
the analysis used to show superintegrability for the TTW system
immediately carries over the the current example and so we obtain a
symmetry operator by raising and lowering the functions associated
with $r$ and $\theta_1$.

Notice that the operators above that raise or lower the
radial eigenfunctions change both $n_0$ and $A_0$.  In order to extend
this
approach and construct a symmetry operator by raising and lowering
the $\theta_1$ and $\theta_2$ functions we will need operators with a
similar
effect on these functions.
This can be achieved using Jacobi function identities
that raise and lower $n$ and $a$ simultaneously when applied to 
\[
\Theta^{(a,b)}_n =
\sin^{a+c}(k\theta)\cos^{b+d}(k\theta)P_{n}^{(a,b)}\cos(2k\theta).
\]
We find
\begin{eqnarray*}
K_n^{+\ a} &=& 
-\frac{(1-a)\cos(k\theta)}{k\sin(k\theta)}\partial_\theta
   - 2(n(n+a+b+1)+a(a+b)) \\
 & & {} \quad - (1-a)(a+c+b+d) - \frac{(1-a)(a-c)}{\sin^2(k\theta)},
\\
K_n^{-\ a} &=& 
-\frac{(1+a)\cos(k\theta)}{k\sin(k\theta)}\partial_\theta
   - 2n(n+a+b+1) \\
 & & {} \quad  - (1+a)(a+c+b+d) + \frac{(1+a)(a+c)}{\sin^2(k\theta)},
\end{eqnarray*}
with action,
\begin{eqnarray*}
 K_n^{+\ a}\Theta^{(a,b)}_n &=& 2(n+1)(n+a)\Theta^{(a-2,b)}_{n+1}, \\
 K_n^{-\ a}\Theta^{(a,b)}_n &=& 2(n+a+b+1)(n+b)\Theta^{(a+2,b)}_{n-1}.
\\
\end{eqnarray*}

\section{Constructing the symmetries}

For $k_1=p_1/q_1$ with gcd$(p_1,q_1)=1$ the operator
\begin{equation}
 \Xi_{1}^+ = \underbrace{K_{0\ n_0-(p_1-1)}^{-\ A_0+2(p_1-1)}\cdots K_{0\ n_0}^{-\ A_0}}_{\mbox{$p_1$ terms}}
    \underbrace{J_{1\ n_1+q_1-1}^{+}\cdots J_{1\ n_1}^+}_{\mbox{$q_1$ terms}}
\label{eqnXi01p} 
\end{equation}
has the effect on a basis function of
\[ n_0\to n_0-p_1, \quad n_1\to n_1+q_1,\quad A_0 \to A_0+2p_1,\]
and so
\begin{eqnarray*}
 E = -2\omega(2n_0 + 2k_1n_1 + \cdots ) 
  &\to& 
 -2\omega(2(n_0-p_1) + 2k_1(n_1+q_1) + \cdots ) \\
 &=& -2\omega(2n_0 + 2k_1n_1 + \cdots ),
\end{eqnarray*}
that is, $E$ is unchanged.  A similar lowering operator is
\begin{equation}
 \Xi_{1}^- = \underbrace{K_{0\ n_0+(p_1-1)}^{+\ A_0-2(p_1-1)}\cdots K_{0\ n_0}^{+\
A_0}}_{\mbox{$p_1$ terms}}
    \underbrace{J_{1\ n_1-(q_1-1)}^{-}\cdots J_{1\ n_1}^-}_{\mbox{$q_1$ terms}},
\label{eqnXi01m} 
\end{equation}
which also leaves $E$ unchanged, has the effect on a basis function of
\[ n_0\to n_0+p_1, \quad n_1\to n_1-q_1,\quad A_0 \to A_0-2p_1.\]
Explicitly, the action of $\Xi_{1}^\pm$ on a basis function is
\begin{eqnarray*}
 \Xi_{1}^+\Psi^{A_0}_{0,n_0}\Psi^{A_1,a_1}_{1,n_1} 
 &=& (-2)^{p_1}\omega^{p_1}(n_1+1)_{q_1}(n_1+A_1+a_1+1)_{q_1}
\Psi^{A_0+2p_1}_{0,n_0-p_1}\Psi^{A_1,a_1}_{1,n_1+q_1},\\
 \Xi_{1}^-\Psi^{A_0}_{0,n_0}\Psi^{A_1,a_1}_{1,n_1} 
 &=& (-2)^{p_1}\omega^{p_1}(-n_1-A_1)_{q_1}(-n_1-a_1)_{q_1}(n_0+p_1)_{p_1}(n_0+A_0)_{p_1}
 \Psi^{A_0-2p_1}_{0,n_0+p_1}\Psi^{A_1,a_1}_{1,n_1-q_1}.
\end{eqnarray*}
This is exactly like the TTW raising and lowering operators from \cite{KKM2011}.

For $k_2/k_1=p_2/q_2$ with gcd$(p_2,q_2)=1$ the operator
\begin{equation}
 \Xi_{2}^+ = \underbrace{K_{1\ n_1-(p_2-1)}^{-\ A_1+2(p_2-1)}\cdots K_{1\ n_1}^{-\ A_1}}_{\mbox{$p_2$ terms}}
    \underbrace{J_{2\ n_2+q_2-1}^{+}\cdots J_{2\ n_2}^+}_{\mbox{$q_2$ terms}}
\label{eqnXi2p}
\end{equation}
has the effect on a basis function of
\[ n_1\to n_1-p_2, \quad n_2\to n_2+q_2,\quad A_1 \to A_1+2p_2\]
and so
\[
 E = -2\omega(2n_0 + 2k_1n_1 + 2k_2n_2 + \cdots ) \to 
 -2\omega(2n_0 + 2k_1(n_1-p_2) + 2k_2(n_2+q_2) + \cdots )
\]
\[ = -2\omega(2n_0 + 2k_1n_1 + 2k_2n_2 + \cdots ),
\]
that is, $E$ is unchanged.  A similar lowering operator is
\begin{equation}
 \Xi_{2}^- = \underbrace{K_{1\ n_1+(p_2-1)}^{+\ A_1-2(p_2-1)}\cdots K_{1\ n_1}^{+\ A_1}}_{\mbox{$p_2$ terms}}
    \underbrace{J_{2\ n_2-(q_2-1)}^{-}\cdots J_{2\ n_2}^-}_{\mbox{$q_2$ terms}}.
\label{eqnXi2m} 
\end{equation}
This is similar to the two-dimensional TTW procedure, but different operators are
required.

The operators given so far are only well defined on the separated basis functions and they
contain
the quantum numbers in their defintions.  We now must show that we can construct pure
differential operators.
The argument is only sketched here as the details are essentially
the same as
those in \cite{KKM2011}.

The transformation $n_1\to-n_1-A_1-a_1-1$ while holding $E$ constant has the
effect of changing the sign of $A_0$.  It is then straightforward to check from the
explicit expressions for the operators that
\[
 L_{1}^+ = \Xi_{1}^+ +  \Xi_{1}^- \qquad\mbox{and}\qquad
 L_{1}^- = k_1\frac{\Xi_{1}^+ -  \Xi_{1}^-}{A_0}
\]
are polynomials in $E$, $A_0^2$ and $A_1^2$.  Since
\[
A_0^2 = k_1^2 - \ell_1 \quad\mbox{and}\quad
A_1^2 = \frac{\frac14k_2^2-\ell_2}{k_1^2}
\]
we can replace $E$, $A_0^2$ and $A_1^2$ with second order differential operators where ever they appear in
these expressions.

Similarly, the transformation $n_2\to-n_2-A_2-a_2-1$ while holding $L_1$ constant has the
effect of changing the sign of $A_1$.  It is then straightforward to check from the
explicit expressions for the operators that
\begin{equation}
\label{eqnhatL2}
 L_{2}^+ = \Xi_{2}^+ +  \Xi_{2}^- \qquad\mbox{and}\qquad
 L_{2}^- = k_2\frac{\Xi_{2}^+ -  \Xi_{2}^-}{A_1}
\end{equation}
are polynomials in $L_1$, $A_1^2$ and $A_2^2$.  Since
\[
A_1^2 = \frac{\frac14k_2^2-\ell_2}{k_1^2} \quad\mbox{and}\quad
A_2^2 = -\frac{\ell_3}{k_2^2}
\]
we can replace $L_1$, $A_1^2$ and $A_2^2$ with a second order differential operators where ever they appear in
these expressions.

In a similar way, we can define $\Xi^\pm_{3}$ and $L^\pm_{3}$ with the label
replacements $1\to2$ and $2\to3$ in (\ref{eqnXi2p}), (\ref{eqnXi2m}) and
(\ref{eqnhatL2}) and
show that they are in also differential operators.

It is clear from the construction that
$\{H,L_1, L^+_{1},L_2, L^+_{2},L_3, L^+_{3}\}$ forms an algebraically
indepenent set
of differential operators and hence the system is superintegrable.

\section{The symmetry algebra}

A common feature of superintegrable systems is a polynomially closed symmetry algebra. 
By direct calcuation, we find some polynomially closed subalgebras of the
symmetry algebra.

Adapting the an argument from \cite{KKM2011} we find that, for $i=1,2,3$,
% \[
%  P^{(+)}_{1}(H,L_1,A_1^2) = \Xi^-_{1}\Xi^+_{1} + \Xi^+_{1}\Xi^-_{1}, \quad
%  P^{(-)}_{1}(H,L_1,A_1^2) = k_1\frac{\Xi^+_{1}\Xi^-_{1} + \Xi^-_{1}\Xi^+_{1}}{A_0},
% \]
% \[
%  P^{(+)}_{2}(L_1,L_2,A_2^2) = \Xi^-_{2}\Xi^+_{2} + \Xi^+_{2}\Xi^-_{2}, \quad
%  P^{(-)}_{2}(L_1,L_2,A_2^2) = k_2\frac{\Xi^+_{2}\Xi^-_{2} + \Xi^-_{2}\Xi^+_{2}}{A_1},
% \]
% \[
%  P^{(+)}_{3}(L_2,L_3,A_3^2) = \Xi^-_{3}\Xi^+_{3} + \Xi^+_{3}\Xi^-_{3}, \quad
%  P^{(-)}_{3}(L_2,L_3,A_3^2) = k_3\frac{\Xi^+_{3}\Xi^-_{3} + \Xi^-_{3}\Xi^+_{3}}{A_2},
% \]
\[
 P^{(+)}_{i}(L_{i-1},L_i,A_i^2) = \Xi^-_{i}\Xi^+_{i} + \Xi^+_{i}\Xi^-_{i}
\quad\mbox{and}\quad
 P^{(-)}_{i}(L_{i-1},L_i,A_i^2) = k_i\frac{\Xi^+_{i}\Xi^-_{i} + \Xi^-_{i}\Xi^+_{i}}{A_{i-1}},
\]
are differential operators that are polynomial in their arguments. 

By comparing the action of brackets
of the operators $\{L_i, L_i^+, L_i^-\}$ with symmetrised products of the operators
we find the following explicit identities for $i=1,2,3$.
\begin{eqnarray*}
 [L_i, L^-_{i}] &=& -4k_i^2q_i^2 L^-_{i} - 4\alpha_i k_i^2q_i L^+_{i} \\
 {} [L_i, L^+_{i}] &=& 2q_i\{L_i, L^-_{i}\} - 4k_i^2q_i L^+_{i} + 4k_i^2q_i^2 L^-_{i} +
8q_i^3k_i^2 L^-_{i} \\
 {} [ L^+_{i}, L^-_{i}] &=& 2q_i( L^-_{i})^2 - 2P^{(-)}_{i}(L_{i-1},L_i,A_i^2)
\end{eqnarray*}
and
\[
 \{L_i, L^-_{i}, L^-_{i}\} + 2k_i^2(14q_i^2-3\alpha_i)( L^-_{i})^2
  +6k_i^2( L^+_{i})^2 + 6k_i^2q_i\{ L^+_{i}, L^-_{i}\} - 12k_i^2P^{(+)}_{i} + 4k_i^2q_iP^{(-)}_{i} = 0,
\]
where $\alpha_1=1$, $\alpha_2=1/4$ and $\alpha_3=0$.  These hold as operator identities
on general functions.
Furthermore, $[L_j,L_i^\pm]=0$ for $i\neq j$ and $[L_j^\pm,L_i^\pm]=[L_j^\pm,L_i^\mp]=0$ for $|i-j|>1$.

% \section{Lower order operators}

As was found for the TTW operators, the symmetries constructed from raising and
lowering operators are not
necessarily of minimal order \cite{KKM2011}.  The same technique for finding lower order
operators can be used for the
current system.

For example, starting from $L_{1}^\pm$ we look for
$M_{1}^\pm$ satisfying
\[
[L_1,M_{1}^\pm]= L_{1}^\pm,\qquad
\]
\[
[H,M_{1}^\pm]=0, \qquad
[L_2,M_{1}^\pm]=0, \qquad
[L_3,M_{1}^\pm]=0. \qquad
\]
Find $M_{1}^-$ is
\begin{equation}
\label{eqnM1m}
 M_{1}^- = -\frac{1}{4q_1}\left( \frac{L_{1}^-}{A_0(A_0+p_1)} + \frac{L_{1}^+}{A_0(A_0-p_1)}\right)
   + \frac{S_1(H,L_2)}{A_0^2-p_1^2}
\end{equation}
where $S_1(H,L_2)$ is a polynomial in $H$ and $L_2$ that can be determined by the methods
used in \cite{KKM2011}.

\section{An example}

Explicit computations can be performed for particular choices of the $k_i$.  For example,
with $k_1,k_2,k_3=2,1,1$, the operator $L_1^-$ is $6$\textsuperscript{th} and $ L_1^+$ is the $5$\textsuperscript{th} order
operator,
\[
L_{1}^+=\left(-\frac2{r^3}\partial_r + \frac6{r^4}\right)A_0^2A_1^2
% \]
%
%
% \[
+\left(-\frac{\cos(4{\theta_1})}{2r^3}{\partial_r}+\frac{\sin(4{\theta_1})}{4r^4}{\partial_{\theta_1}}
+\frac{1+5\cos(4{\theta_1})}{2r^4}\right){A_0}^4
\]
\[
+\left(-\frac1r{\partial_r}+\frac2{r^2}\right){E}{A_1}^2
% \]
%
%
% \[
+\left(\frac{\sin(4{\theta_1})}{16}{\partial_{\theta_1}}
+\frac{\cos(4{\theta_1})}{4}+\frac18\right){E}^2
\]
\[
+\left(
-\frac{\cos(4{\theta_1})}{4r}{\partial_r}
+\frac{\sin(4{\theta_1})}{4r^2}{\partial_{\theta_1}}
+\frac{3\cos(4{\theta_1})+1}{2r^2}
\right)E{A_0}^2
% \]
%
%
% \[
+\left(-\frac{10}{r^3}{\partial_r}+\frac{4}{r^2}{\partial_{r}^2}-\frac{6}{r^4}\right){A_1}^2
\]
\[
-\left(
\frac{\sin(4{\theta_1})}{r}\partial_r\partial_{\theta_1}
+\frac{3\cos(4{\theta_1})+2-a_1^2}{r}{\partial_r}
-\frac{5\sin(4{\theta_1})}{4r^2}{\partial_{\theta_1}}
-\frac{(6\cos(4{\theta_1})+5-4a_1^2)}{2r^2}
\right)E
\]
\[
+\left(
-\frac{\sin(4{\theta_1})}{4r^2}{\partial_{r}^2}{\partial_{\theta_1}}
+\frac{13\sin(4{\theta_1})}{4r^3}\partial_r\partial_{\theta_1}
-\frac{(4\cos(4{\theta_1})+1)}{2r^2}{\partial_{r}^2}
+\frac{13+27\cos(4{\theta_1})-4a_1^2}{2r^3}{\partial_r}
\right.
\]
\[
\left.
-\frac{5\sin(4{\theta_1})}{r^4}{\partial_{\theta_1}}
-\frac{25\cos(4{\theta_1})+20-12a_1^2}{2r^4}
\right){A_0}^2
\]
\[
+\frac{11\sin(4{\theta_1})}{4r^2}{\partial_{r}^2}{\partial_{\theta_1}}
+\frac{(11-8a_1^2+14\cos(4{\theta_1}))}{2r^2}{\partial_{r}^2}
-\frac{23\sin(4{\theta_1})}{4r^3}\partial_r\partial_{\theta_1}
\]
\[
-\frac{26\cos(4{\theta_1})+23-20a_1^2}{2r^3}{\partial_r}
-\frac{21\sin(4{\theta_1})}{4r^4}{\partial_{\theta_1}}
-\frac{3(10\cos(4{\theta_1})+7-4a_1^2)}{2r^4}
\]
with the replacements \ $E \to H$,
$A_0^2 \to k_1^2 - L_1$ and
$A_1^2 \to (k_2^2 - 4L_2)/(4k_1^2)$.

A 4\textsuperscript{th} order operator can also be constructed using
(\ref{eqnM1m}) and in this
case,
\[
S_1(H,L_2) =  -\frac{(H^2-4\omega)(A_1^2-a_1^2)}{16}
\]
and $M_{1}^+$ is a polynomial in $E$ and even powers of $A_0$ and $A_1$.

\section{Conclusion}

The methods developed in \cite{KKM2011} have been extended to
demonstrate the superintegrability of a 4-dimensional quantum Hamiltonian system
on a non-conformally-flat space.  The system discussed is a quantisation of a previously
described classical superintegrable system and in order to maintain superintegrability in
the quantisation, correction terms were required to be added to the potential.  These
correction terms make the Hamiltonian
conformally covariant, but are not the usual minimal conformally covariant
correction of $-\mathcal{R}/6$ as they depend on the conformal curvature.  The
3-dimensional analogue of this system also requires the addition of the term
$-\mathcal{R}/8$ to maintain superintegrability and this too gives a conformally
covariant Hamiltonian.

While many previously known superintegrable systems possess a polynomially closed symmetry
algebra, here
we have only found some polynomially closed subalgebras.  An investigation of a
closely related classical system found that in general the symmetry algebra will close
rationally rather than polynomially \cite{KM2012}.  It seems reasonable to conjecture
that,
except in some special cases, the symmetry algebra of the 4-dimensional system
considered here does not close polynomially, but rather
it obeys an appropriate quantum analogue of rational closure.

\subsection*{Acknowledgement}
This work was partially supported by a grant from the Simons Foundation (\# 208754 to
Willard Miller, Jr.).

\label{thelastpage}

\end{document}